# AIRPORT CYBER SECURITY & CYBER RESILIENCE CONTROLS

Alex Mathew[1]

[1]Department of Computer Science & Cyber Security, Bethany College, WV, USA.
amathew@bethanywv.edu

**ABSTRACT**

*Cyber Security scares are the main areas of demerits associated with the advent and widespread of internet technology. While the internet has improved life and business processes, the levels of security threats have been increasing proportionally. As such, the web and the related cyber systems have exposed the world to the state of continuous vigilance because of the existential threats of attacks. Criminals are in the constant state of attempting cybersecurity defense of various infrastructures and businesses. Airports are some of the areas where cybersecurity means a lot of things. The reason for the criticality of cybersecurity in airports concerns the high integration of internet and computer systems in the operations of airports. This paper is about airport cybersecurity and resilience controls. At the start of the article is a comprehensive introduction that provides a preview of the entire content. In the paper, there are discussions of airport intelligence classification, cybersecurity malicious threats analysis, and research methodology. A concise conclusion marks the end of the article.*

## 1. INTRODUCTION

Since the 9/11 incident and the increasing interests of terrorists in aircraft and points of entry and departures, airports have become critical security areas. After 9/11, many people internalized airport and flight security as serious matters. The global community has increased the level of airport and aircraft vigilances. Besides the physical threats, cyber systems of airports have become critical targets for attackers. As such, airports have had to increase their levels of cybersecurity alertness and resilience to prevent breaches. Internet of things (IoT)

Internet of things (IoT) is the technical architecture that involves the interconnectivity of smart devices and tools that communicate regularly. Almost every electronic manufactured today has capabilities for connecting to the IoT architecture [10]. IoT has become a necessary technology in airports, as it helps in facilitating communication among various intelligent devices and systems. IoT solutions in airports have created the opportunities for visualization of safety, hyper-personalized experience for customers, and efficient operations [4]. IoTs have built the most revolutionary airports. IoTs in airports expedite airport operations through smart sensors, analytics, and connected devices. IoT has helped in enhancing cyber resilience and operational efficiencies. Airports have benefitted from the application of electronic sensors and tags to trace and track baggage and addressing the problem of luggage thefts. Facial recognition innovations streamline security checks and cut queues. The integration of IoT in airports has also created new opportunities for revenue generation [9]. Despite the various benefits of IoT integration in airports, the interconnectivity of multiple devices and systems complicate cybersecurity because of many points of possible attacks.

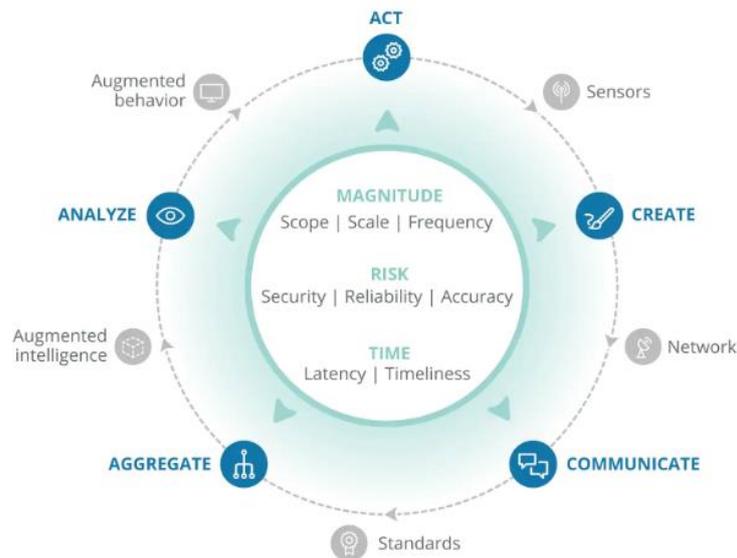

**Figure 1: Information value loop describing the architecture of IoT**

Image retrieved from [6]

## II. AIRPORT INTELLIGENCE CLASSIFICATION

Airport intelligence classification attributed to good technical practices is a high level of security issue. The good technical practice involves cautious and preventive measures like the installation and regular updating of anti-malware software. It also consists of the education and training of employees on responsible behaviors while interacting with IT resources [10]. Such training and education can involve accountable password management and communication with outsiders. Regular data backup also constitutes an excellent technical practice. The culture of regular data backup safeguards airports from total closure in the event of data and system attack.

Policies and safety standards aspect of airport intelligence classification is an essential security issue that concerns the authoring and communication of clear policies and safety measures. Good practices require that organizations develop clear and attainable strategies as well as safety regulations that reflect the current and future cybersecurity needs [7]. Wrong policies and safety standards or poor communication of the rules threaten the stability of a cyber-system and exposes it to attacks.

The proper organizational practice of airport intelligence classification concerns the efforts made by individual institutions to enhance cybersecurity resilience. Such efforts include investment plans in cybersecurity tools and structures. The proper organizational practice also involves partnering with other industry actors and regulatory authorities to explore potential threats and preventive strategies [1]. The implementation of industry standards is another aspect of good organizational practice. Organizations that are sensitive about cyber threats ought to implement the set industry standards designed to enhance resilience against breaches.

### III. CYBERSECURITY MALICIOUS THREATS ANALYSIS

Airports are susceptible to various malicious threats that can disrupt and compromise operations. Malicious threats are a category of cybersecurity scares that anti-virus cannot tackle. Most common malicious threats include sniffing, spear phishing, execution of remote codes, pharming, man-in-the-middle, and denial-of-Service-DoS [7]. The increasing integration of the internet of things (IoT) in the airports is increasing the vulnerabilities of networks to attacks. The malicious threats that face smartest airport because of the IoT infrastructure divide into "malicious software, misuse of authorization, social and phishing attacks, network and communication attacks, and tampering with airport smart devices" [5].

### IV. RESEARCH METHODOLOGY

The study on airport cybersecurity and cyber resilience will employ qualitative research approach. Various reasons inform the choice for the qualitative technique. One of the reasons is that the qualitative approach allows for a detailed evaluation of materials to glean more in-depth information on a given subject or topic [8]. The exploration of cybersecurity in airports cannot be substantive and detailed with other methods aside from qualitative techniques. Another benefit inspiring the choice of qualitative technique attributes to its reliance on human observations and experiences. Airport IT personnel are the best positioned to comment on their experiences with cybersecurity issues. The testimonies by IT personnel in airports can give a more in-depth view of the situation that can be instrumental in predicting the future of cyber-security [4]. Finally, the data collected through qualitative means have predictive qualities. Researchers can observe the trend of responses and deduce predictions about a situation.

The preferred data collection approaches will be interview and document review. The researchers will interview IT officials of various airports. The reason for choosing the interview method attributes to its ability to produce detailed facts and views about the subject of airport cybersecurity — another purpose for selecting interviewing concerns time-saving capability. Interviews are easy to conduct and consume the least time [3]. The originality of the data gathered through interviews is another reason for its preference to other techniques of data collection. Besides the interviews, document analysis provides access to a wide variety of data. The review of documents can offer researchers the opportunity to access information that is difficult to find through interviews [2]. The data collected through document analysis also provide materials for comparison with the responses of interviewees. As such, it becomes possible to verify every information.

The methods of data recording will include note-taking, audio recording, and videos. All three ways have the shared benefit of cost-effectiveness and long life. Notes, adios, and videos require a small budget for a pen, notebook, and a recording device that be researchers' smartphone. The data can also last for a long time and serve as reference points for future needs [2].

## V. CONCLUSION

Cyber and internet control most of the life activities in the contemporary world. The situation means that the world is susceptible to the consequences of the increasing dependency on the internet and related systems. Cyber threats and attacks are the main dangerous consequences of dependence on the internet. The attacks and threats occur in multiple ways. All sectors using the cyber system and the internet to run operation have experienced some forms of attacks and threats. The aviation sector has had various cybersecurity scare on airports. It is worth noting that airports are increasingly become smart because of the enhanced connectivity of systems and processes. Internet of things (IoT) is the best way to describe the connectivity of devices in modern airports. This sophisticated connectivity of devices and systems has created many loopholes that have exposed airports to cyber-attacks. While some airports have suffered attacks, the industry is continually evolving and improving its cybersecurity strategies.

**Authors**

**Alex Mathew** Ph.D., CISSP, CEH, CHFI, ECSA, MCSE, CCNA. Security+

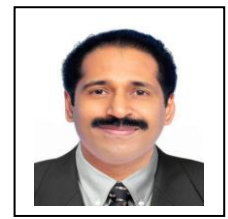